\documentclass{elsart}
\usepackage{graphicx}
\usepackage{dcolumn}
\usepackage{bm}

\begin{document}
\begin{frontmatter}
\title{Isotopic dependence of the pygmy dipole resonance}
\author{N. Paar}
\address{Institut f\" ur Kernphysik, Technische Universit\" at Darmstadt Schlossgartenstr. 9,
D-64289 Darmstadt, Germany}
\author{T. Nik\v si\' c and D. Vretenar}
\address{Physics Department, Faculty of Science, University of Zagreb, Croatia, and \\
Physik-Department der Technischen Universit\"at M\"unchen, D-85748 Garching,
Germany}
\author{P. Ring}
\address{Physik-Department der Technischen Universit\"at M\"unchen, D-85748 Garching,
Germany}
\date{\today}

\begin{abstract}
The isotopic dependence of the excitation energies of the pygmy 
dipole resonance (PDR) is analyzed in the framework of the self-consistent
relativistic Hartree-Bogoliubov (RHB) model and the 
relativistic quasiparticle random-phase approximation (RQRPA).
The DD-ME1 density-dependent meson-exchange 
interaction is used in the effective 
mean-field Lagrangian, and pairing correlations are described
by the pairing part of the finite-range Gogny interaction D1S. 
Model calculations reproduce available experimental data on charge radii, 
the neutron skin, neutron separation energies, and excitation energies
of isovector giant dipole resonances in Ni, Sn and Pb nuclei.
In all three isotopic chains the one-neutron separation energies
decrease with mass number much faster than the excitation energies
of the PDR. As a result, already at moderate proton-neutron asymmetry
the PDR peak energy is calculated above the neutron emission threshold.
This result has important implications for the observation of the PDR in 
$(\gamma,\gamma^\prime)$ experiments. 

\end{abstract}
\end{frontmatter}

Studies of the structure and stability of nuclei with extreme isospin
values provide new insights into every aspect of the nuclear many-body
problem. In neutron-rich nuclei far from the valley of $\beta$-stability,
in particular, new shell structures occur 
as a result of the modification of the effective 
nuclear potential. Neutron density distributions become very diffuse and
the phenomenon of the evolution of the neutron skin and, in some cases,
the neutron halo have been observed. 
The weak binding of outermost neutrons gives rise to soft excitation 
modes. In particular, the pygmy dipole resonance (PDR), i.e. the resonant
oscillation of the weakly-bound neutron mantle against the isospin
saturated proton-neutron core, has recently been the subject of a number of
theoretical and experimental studies.  Its structure, however, remains
very much under discussion. Properties of the PDR in neutron-rich nuclei
have important implications on theoretical predictions
of the radiative neutron capture rates
in the r-process nucleosynthesis, and consequently to the calculated  
elemental abundance distribution~\cite{Gor.98,Gor.02,Gor.04}. 
Furthermore, the detailed knowledge of the structure of low-energy modes
of excitation would also place stringent constraints on the isovector
channel of effective nuclear interactions.
An interesting problem is the isotopic dependence of the PDR, and 
especially the behavior of the PDR 
in the vicinity of major spherical shell gaps.
Already in a recent study of low-lying $2^+$ excitations
in the region of the N=82 magic neutron number~\cite{Rad.02}, it has been 
observed that the lowering of the first excited $2^+$ state in neutron-rich 
Te isotopes is accompanied by a reduction of the 
corresponding E2 strength, in contradiction with systematics 
and general expectations about quadrupole collectivity. 
The anomalous behavior of the Te isotopes has been explained by the 
weakening of neutron pairing above the N=82 spherical gap~\cite{Ter.02}.

A systematic analysis of the dipole response of light exotic
nuclei has been recently reported in the experimental study of electromagnetic 
excitations of oxygen isotopes in heavy-ion collisions~\cite{Lei.01}. 
For neutron-rich oxygen isotopes
the resulting photo-neutron cross sections  are characterized
by a pronounced concentration of low-lying E1 strength.  
The onset of low-lying E1 strength has been observed not only 
in exotic nuclei with a large neutron excess,
but also in stable nuclei with moderate proton-neutron asymmetry. 
High resolution photon scattering has been employed to measure 
low-lying dipole strength distributions in 
Ca~\cite{Har.00}, Sn~\cite{Gov.98} and Pb~\cite{Rye.02,End.03} isotopes,
and in the N=82~\cite{Zil.02} isotone chain. 

A number of theoretical approaches have been employed in the investigation
of the nature of the low-lying dipole strength: the Steinwedel - 
Jensen hydrodynamical model~\cite{Suz.90}, density
functional theory~\cite{Cha.94}, large-scale shell-model
calculations~\cite{Sag.99},
the self-consistent Skyrme Hartree-Fock + RPA model~\cite{Rei.99},
Skyrme Hartree-Fock + QRPA with phonon coupling~\cite{Col.01},
time-dependent density-matrix theory~\cite{Toh.01}, continuum
linear response in the coordinate-space Hartree-Fock Bogoliubov
formalism~\cite{Mat.01}, the quasiparticle phonon
model~\cite{Rye.02,End.03,TLS.04,TLS.04a}, the relativistic 
RPA~\cite{Vrepyg2.01}, and the relativistic QRPA in the
canonical basis of the Relativistic Hartree-Bogoliubov 
model~\cite{Paar.03}. In general, 
the dipole response of very neutron-rich isotopes is characterized by the
fragmentation of the strength distribution and its spreading into the
low-energy region, and by the mixing of isoscalar and isovector modes. 
In relatively light nuclei the onset of dipole strength in
the low-energy region is due to non-resonant independent single particle
excitations of the loosely bound neutrons. 
However, the structure of the low-lying
dipole strength changes with mass. As we have shown in the RRPA analysis of
Ref.~\cite{Vrepyg2.01}, in heavier nuclei low-lying dipole states appear
that are characterized by a more distributed structure of the RRPA
amplitude. Among several peaks characterized by single particle transitions,
a single collective dipole state is identified below 10 MeV, and its
amplitude represents a coherent superposition of many neutron particle-hole
configurations.

In this work we use the relativistic quasiparticle random phase 
approximation (RQRPA) based on the canonical 
single-nucleon basis of the relativistic
Hartree-Bogoliubov (RHB) model, to
analyze the isotopic dependence of dipole pygmy resonances 
in medium-heavy and heavy nuclei.
The RHB model provides a unified description of particle-hole 
($ph$) and particle-particle ($pp$) correlations. 
In the canonical basis, in particular, the ground state 
of a nucleus takes the form of a highly correlated BCS-state. 
By definition, the canonical basis diagonalizes the density matrix and 
is always localized. It describes both the bound states and the 
positive-energy single-particle
continuum. The formulation of the RQRPA in the canonical basis is
particularly convenient because, in order to describe transitions to
low-lying excited states in weakly bound nuclei, the
two-quasiparticle configuration space must include states with both nucleons
in the discrete bound levels, states with one nucleon in a bound level and
one nucleon in the continuum, and also states with both nucleons in the
continuum.

The relativistic QRPA of Ref.~\cite{Paar.03} is fully self-consistent.
For the interaction in the particle-hole channel effective Lagrangians with
nonlinear meson self-interactions or density-dependent 
meson-nucleon couplings are used, and pairing correlations are
described by the pairing part of the finite-range Gogny interaction. Both in
the $ph$ and $pp$ channels, the same interactions are used in the RHB
equations that determine the canonical quasiparticle basis, and in the
matrix equations of the RQRPA. This feature is essential for the
decoupling of the zero-energy mode which corresponds to the spurious
center-of-mass motion. In addition to configurations built from
two-quasiparticle states of positive energy, the RQRPA configuration space
contains quasiparticle excitations formed from the 
ground-state configurations of fully or partially occupied
states of positive energy and the empty negative-energy states from the
Dirac sea. 
 
The R(Q)RPA model has
been successfully employed in analyses of low-lying quadrupole and dipole 
states~\cite{Vrepyg2.01,Paar.03}, multipole giant 
resonances~\cite{Ma.01},
toroidal dipole resonances~\cite{Vre.02}, isobaric analog
and Gamow-Teller resonances~\cite{Vre.03}. 
In Ref.~\cite{Nik2.02} we have extended the RRPA framework to include 
relativistic effective mean-field interactions with 
density-dependent meson-nucleon couplings. In a number of 
recent studies it has been shown that, in comparison 
with standard RMF effective interactions with nonlinear meson-exchange
terms, density-dependent meson-nucleon interactions
significantly improve the description of asymmetric nuclear matter 
and isovector ground-state properties of finite nuclei. This is,
of course, very important for the extension of RMF-based models 
to exotic nuclei far from $\beta$-stability.
In particular, one expects that the properties of pygmy modes
will be closely related to the size of the neutron skin~\cite{TLS.04}
and, therefore, in a quantitative analysis it is important to 
use effective interactions that reproduce available data on 
the neutron skin. 
 
\begin{figure}
\includegraphics[scale=0.6,angle=0]{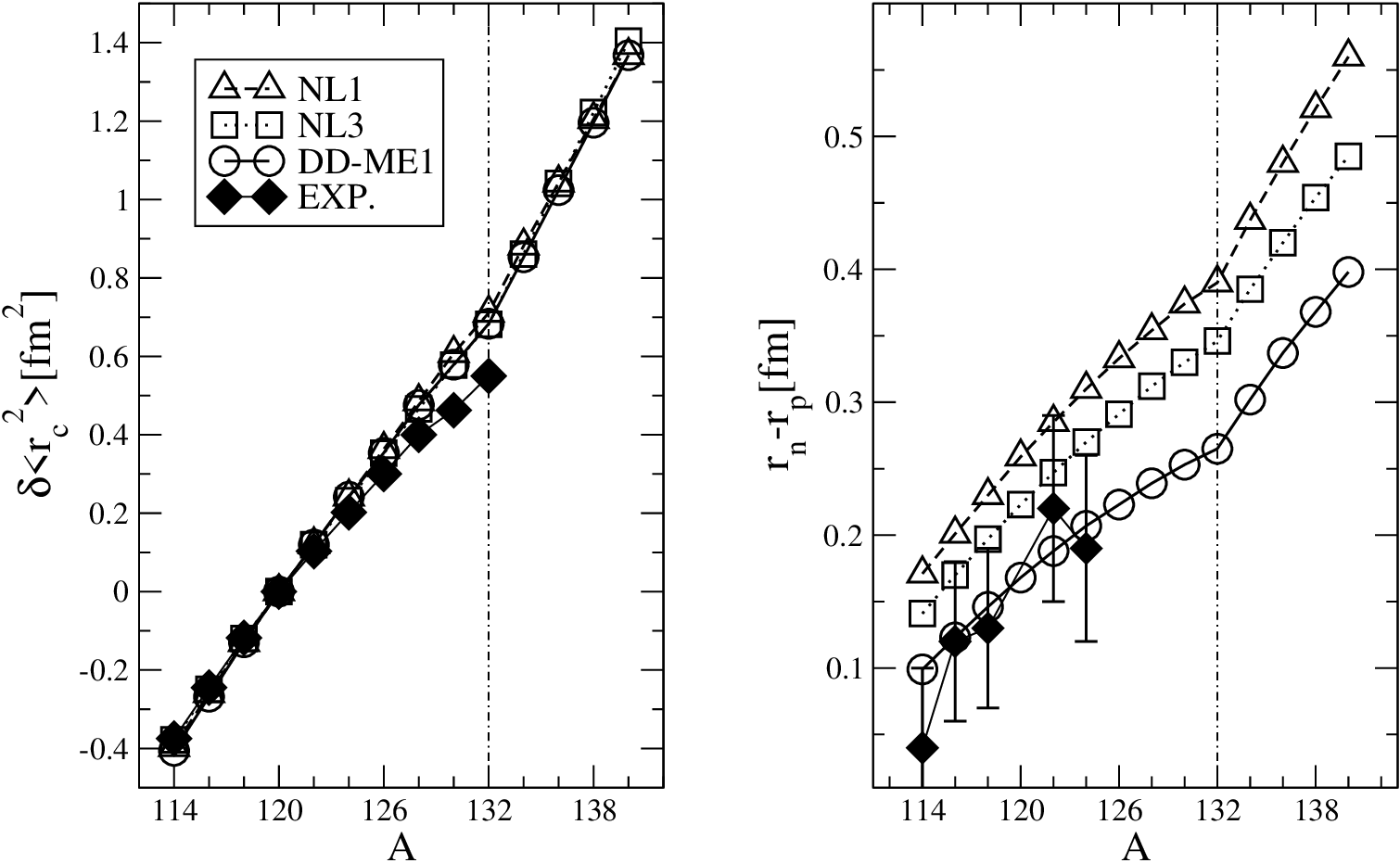}
\caption{Charge isotope shifts (left panel), and 
the differences between the radii of neutron and proton
density distributions (right panel) for the Sn isotopes,
as functions of the mass number.  
The values calculated in the RHB model with the 
NL1, NL3 and DD-ME1 effective
interactions, are shown in comparison with 
empirical data~\protect\cite{CHARGE,Blanc.02,Kra.99}.}
\label{fig1}
\end{figure}

In the present study the density-dependent effective 
interaction DD-ME1~\cite{Nik1.02} is employed in the RQRPA calculation 
of the dipole response in the Sn, Ni and Pb isotopic chains. 
In Fig.~\ref{fig1} we display the self-consistent RHB model results for 
the charge isotope shifts and the differences between the radii 
of the neutron and proton density distributions of Sn isotopes.
The Gogny interaction D1S~\cite{BGG.84} is used in the pairing channel, and 
the radii calculated with the DD-ME1 interaction, and with two 
standard non-linear effective interactions NL1~\cite{RRM.86} and
NL3~\cite{LKR.97}, are shown in comparison with available empirical 
data~\cite{CHARGE,Blanc.02,Kra.99}. Although the charge radii 
calculated with all three effective interactions are in very good  
agreement with experimental data, only DD-ME1 quantitatively 
reproduces the evolution of the neutron skin. 
Because of their high asymmetry energy at saturation density, 
NL1 and NL3 predict much larger neutron radii. A high value of 
the asymmetry energy characterizes all the standard RMF forces
with non-linear meson-exchange terms and, therefore, large neutron 
radii are obtained with any of these interactions. In the 
following analysis we employ the RHB+RQRPA model  
with the density-dependent interaction DD-ME1 in the
$ph$-channel, and with the finite range Gogny interaction D1S in
the $pp$-channel.

\begin{figure}
\includegraphics[scale=0.6,angle=0]{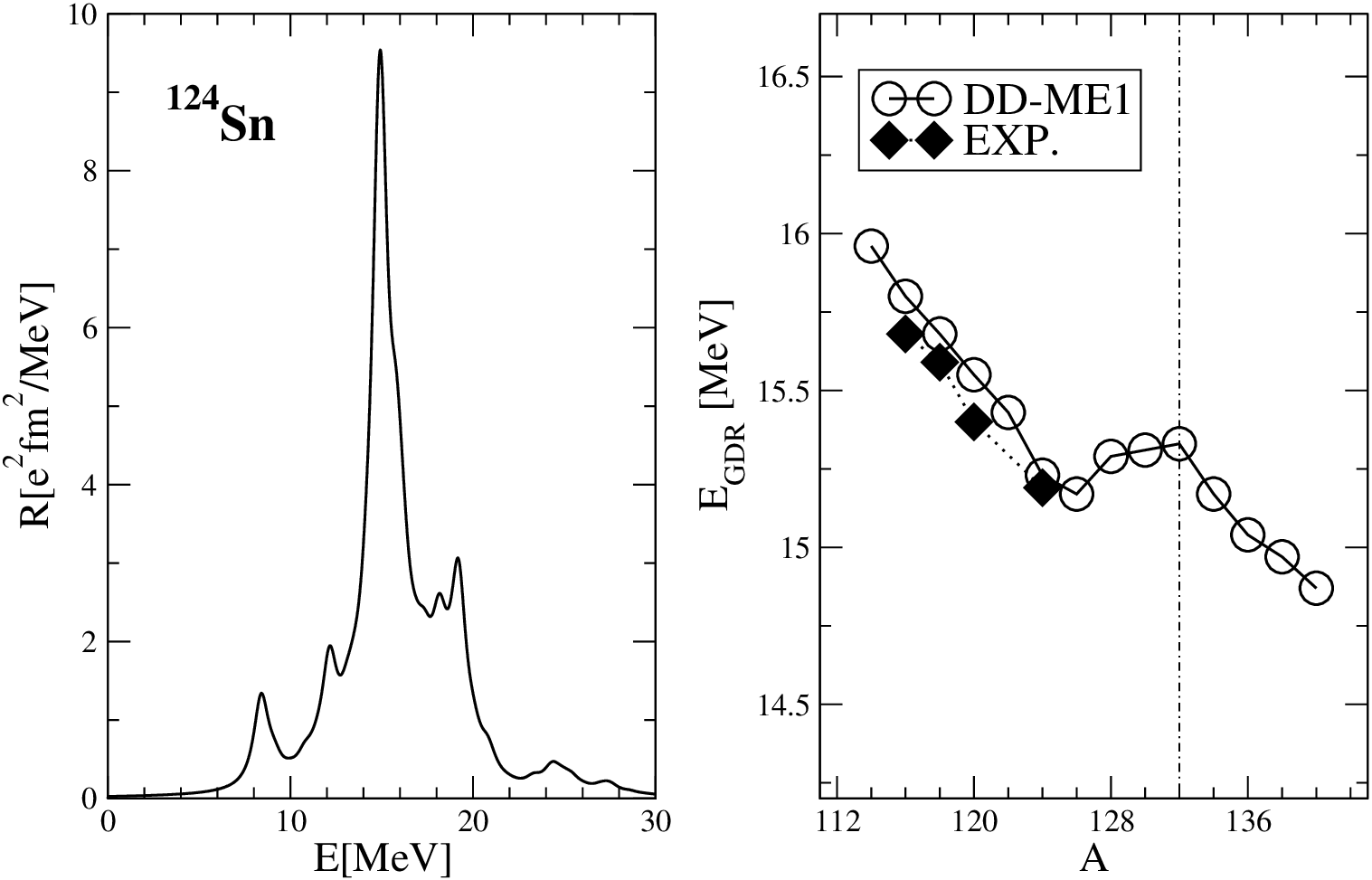}
\caption{The RHB+RQRPA isovector dipole strength distribution 
in $^{124}$Sn (left panel). The experimental IV GDR excitation
energies for the Sn isotopes are compared with the RHB+RQRPA
results calculated with the DD-ME1 effective interaction (right panel).}
\label{fig2}
\end{figure}

In Fig.~\ref{fig2} we display the isovector dipole strength
distribution in $^{124}$Sn. The calculation is fully self-consistent, 
with the Gogny finite-range pairing
included both in the RHB ground state, and in the RQRPA residual
interaction. In addition to the characteristic
peak of the isovector giant dipole resonance (IVGDR) at $\approx$ 15 MeV,
among several dipole states in the low-energy
region between 7 MeV and 10 MeV that are characterized by single particle
transitions, at $\approx$ 8.5 MeV a single pronounced peak is found  
with a more distributed structure of
the RQRPA amplitude, exhausting 3.3\% of the TRK sum rule.
As we have shown in Ref.~\cite{Paar.03} by 
analyzing the corresponding transition 
densities, the dynamics of this low-energy mode is very different from that
of the IVGDR: the proton and neutron transition densities
are in phase in the nuclear interior, 
there is almost no contribution from the protons
in the surface region, the isoscalar transition density dominates
over the isovector one in the interior, and  
the large neutron component in the surface
region contributes to the formation of a node in the isoscalar 
transition density. The low-lying pygmy 
dipole resonance (PDR) does not belong to statistical E1 
excitations sitting on the tail of the GDR, but represents a 
fundamental structure effect: the neutron skin oscillates against the core.
In the right panel of Fig.~\ref{fig2} we 
compare the RQRPA results for the Sn isotopes with
experimental data on IVGDR excitation energies~\cite{Ber.75}. 
The energy of the resonance $E_{\mathrm{GDR}}$ is defined as the centroid 
energy $\bar E = {m_1}/{m_0}$, calculated in the same 
energy window as the one used in the experimental analysis (13--18 MeV). 
The GDR excitation energies decrease with 
increasing mass number, but the mass dependence is not 
monotonic. The calculation predicts an increase 
of the GDR excitation energy at the $N=82$ shell closure. We notice that
for the four isotopes for which data have been reported,
the calculated energies of the GDR are in excellent agreement with
the experimental values.  

\begin{figure}
\includegraphics[scale=0.5,angle=0]{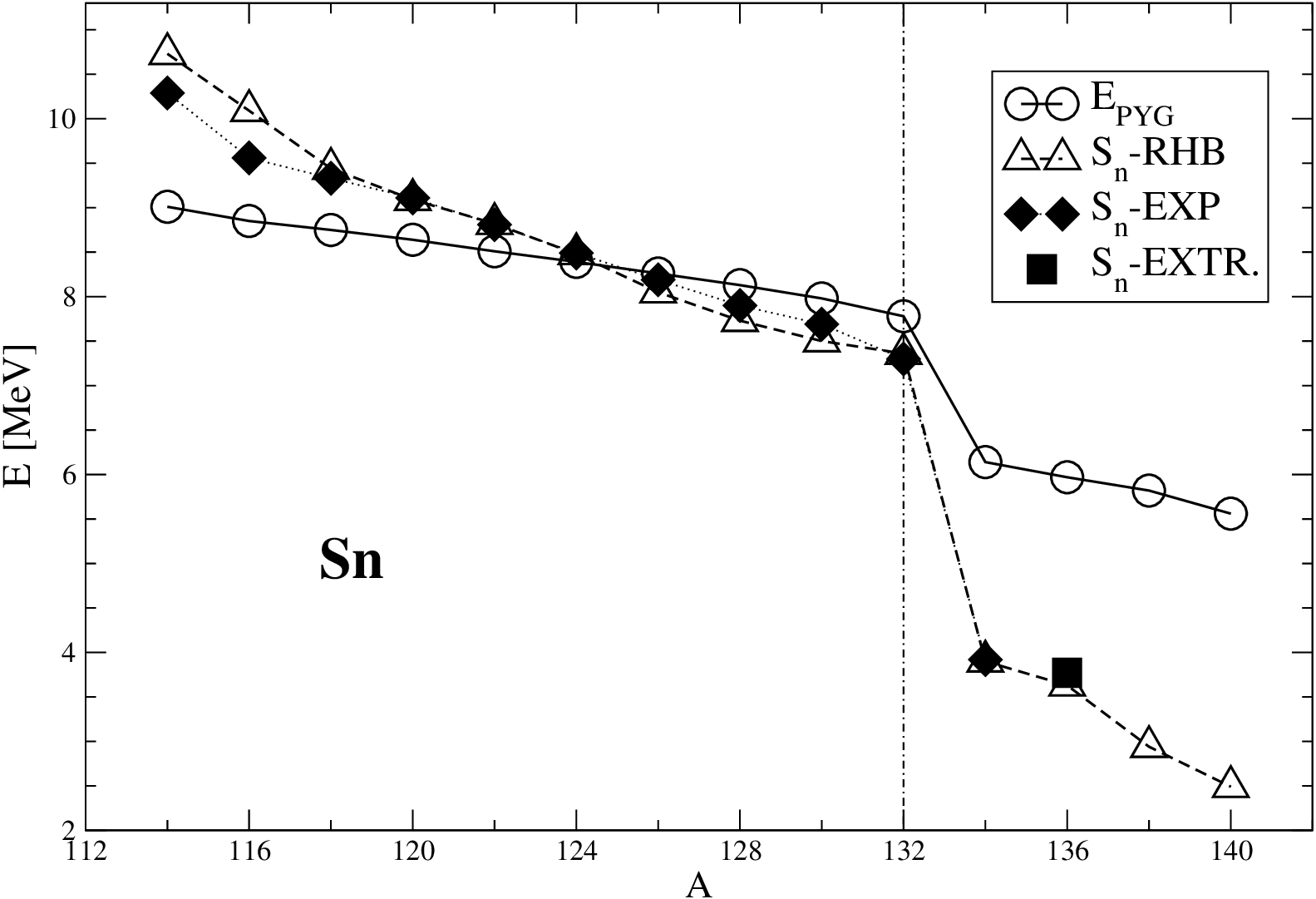}
\caption{The calculated PDR peak energies and the 
one-neutron separation energies for the sequence of Sn isotopes, as 
functions of the mass number. The DD-ME1 effective interaction 
has been used in the RHB and RQRPA calculations. The RHB results for the 
neutron separation energies are compared with the 
experimental and extrapolated values~\protect\cite{AW.95}.
}
\label{fig3}
\end{figure}

In Fig.~\ref{fig3} the calculated peak energies of the PDR in Sn isotopes 
are plotted as function of the mass number. The RQRPA predicts a
monotonic decrease of the PDR with mass number, and only a small 
kink in the calculated excitation energies is found at the 
$N=82$ shell closure. In the same plot we have also included 
the calculated one-neutron separation energies, in comparison
with the experimental data and the extrapolated value from the 
compilation of Audi and Wapstra~\cite{AW.95}. The self-consistent 
RHB calculation, with the DD-ME1 mean-field effective interaction 
and the D1S Gogny pairing force, reproduces in detail the one-neutron 
separation energies of the Sn nuclei. We notice
that the separation energies decrease much faster than the 
calculated PDR excitation energies. At $N=82$, in 
particular, the separation energies display a sharp decrease, whereas
the shell closure produces only a weak effect on the PDR excitation 
energies. The important result here is that for $A < 124$ the PDR 
excitation energies are lower than the corresponding one-neutron 
separation energies, whereas for $A\geq 124$ the pygmy resonance
is located above the neutron emission threshold. This means, of course,
that in the latter case the observation of the PDR in $(\gamma,\gamma^\prime)$ 
experiments will be strongly hindered. One would naively expect that, in
a given isotopic chain, the relative strength of 
the PDR increases monotonically with the number of neutrons, 
at least within a major shell. In Ref.~\cite{Paar.03} we have shown, however, 
that in the case of Sn isotopes the PDR peak is most pronounced in 
$^{124}$Sn. A combination of shell effects and reduced 
pairing correlations, decrease the relative strength of the PDR in heavier
Sn nuclei below $N=82$.  

\begin{figure}
\includegraphics[scale=0.6,angle=0]{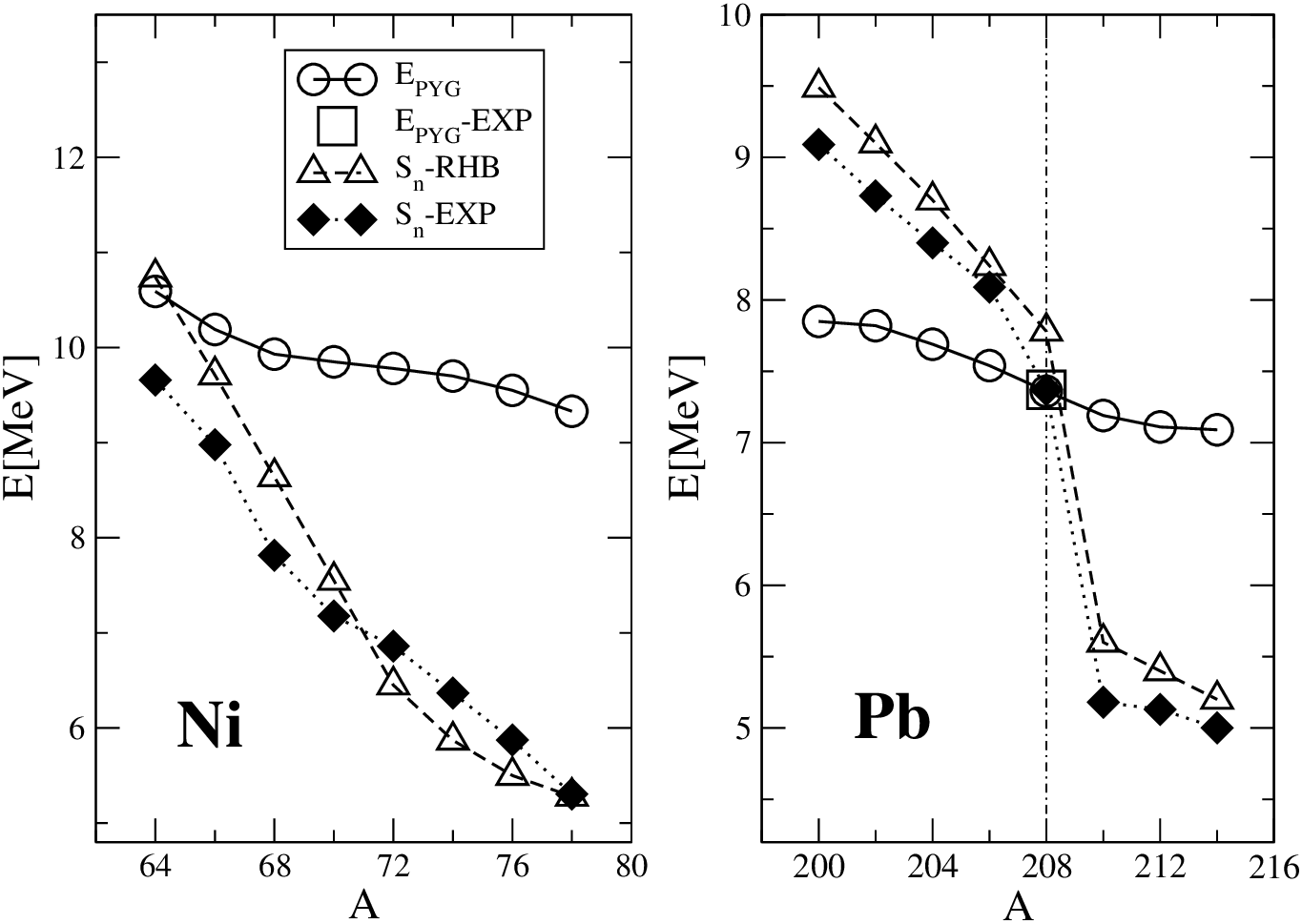}
\caption{Same as in Fig.~\protect\ref{fig3}, but for the Ni and 
Pb isotopic chains. The open square denotes the experimental position 
of the PDR in $^{208}$Pb~\protect\cite{Rye.02}.
}
\label{fig4}
\end{figure}

Besides being intrinsically interesting as an exotic mode of excitation, 
the occurrence of the PDR might have important implications for the 
r-process nucleosynthesis. Namely, although the E1 strength of 
the PDR is small compared to the total dipole strength, if located 
well below the neutron separation energy the PDR can significantly
enhance the radiative neutron capture cross section on neutron-rich
nuclei, as shown in recent large-scale QRPA calculations 
of the E1 strength for the whole nuclear chart \cite{Gor.02,Gor.04}.
The results of the present RHB+QRPA calculation
show, however, that in very neutron-rich nuclei the PDR could be 
located well above the neutron emission threshold (see also 
Fig.~\ref{fig4}). 

It is, of course, interesting to explore other isotopic chains of 
spherical nuclei where one expects, or has already observed, 
the occurrence of the PDR in the E1 excitation spectrum. In Fig.~\ref{fig4} 
we plot the calculated PDR excitation energies and the 
one-neutron separation energies for the Ni and Pb isotopic chains. 
As in the case of Sn isotopes, in Pb nuclei the PDR strength is always
concentrated in one peak and we display the corresponding peak energies.
On the other hand Ni nuclei are much lighter and the 
PDR strength is fragmented over several states. Therefore the 
excitation energies shown in the left panel of Fig.~\ref{fig4} 
correspond to the centroid energies of the PDR strength distributions. 
The RHB results for the neutron separation energies are
compared with the experimental values~\cite{AW.95}. For both chains
the RQRPA calculation predicts a very weak mass dependence of the PDR
excitation energies. In the sequence of Ni isotopes the crossing between 
the theoretical curves of one-neutron separation energies and PDR 
excitation energies is calculated already at $A=64$. In heavier, neutron-rich 
Ni nuclei the PDR is expected to be located high above the neutron 
emission threshold. Notice that for the Ni isotopic 
chain the agreement between the calculated and experimental neutron 
separation energies is not as good as for the Sn nuclei and, therefore, 
the actual point of crossing between the PDR and the one-neutron separation 
energy could occur for $A<64$. The Ni isotopes are not very rigid nuclei and, 
for a more quantitative prediction, one would probably have to go beyond the 
simple mean-field plus QRPA calculation and include correlation effects.
For the Pb isotopes the crossing point is
calculated at $A=208$, in excellent agreement with very recent experimental
data on the PDR in $^{208}$Pb~\cite{Rye.02}. 
Future $(\gamma,\gamma^\prime)$ experiments on Pb nuclei could confirm
the predictions of the present analysis. 

In conclusion, we have employed the self-consistent RHB model and the 
RQRPA in the analysis of the isotopic dependence of the excitation 
energies of the PDR in the Ni, Sn, and Pb isotopic chains. 
By using the density-dependent DD-ME1 meson-exchange effective 
interaction, which reproduces available data on the charge radii, 
the neutron skin, and neutron separation energies, we have 
shown that the RQRPA results for the excitation energies of 
the isovector giant dipole resonances are in excellent agreement
with data in Sn nuclei. We have then compared the isotopic dependence 
of the excitation energies of the PDR with that of the corresponding 
one-neutron separation energies. The separation energies decrease
much faster with mass number. Especially at shell closures the 
one-neutron separation energies display a sharp decrease, whereas
only a weak discontinuity is predicted for the PDR peak energies. 
In all three isotopic chains the RQRPA predicts a crossing between
the curves of neutron separation energies and PDR excitation energies 
already at moderate proton-neutron asymmetry. For the heavier 
isotopes the PDR is calculated above the 
neutron emission threshold, and this will effectively 
preclude the observation of the PDR in $(\gamma,\gamma^\prime)$ 
experiments on very neutron-rich nuclei. 

\bigskip 
\leftline{\bf ACKNOWLEDGMENTS}

This work has been supported in part by the Bundesministerium
f\"ur Bildung und Forschung under project 06 MT 193, and by the
Gesellschaft f\" ur Schwerionenforschung (GSI) Darmstadt.
N.P. acknowledges support from the Deutsche
Forschungsgemeinschaft (DFG) under contract SFB 634.

\bigskip \bigskip

\newpage

\begin{thebibliography}{999}
\bibitem{Gor.98} S. Goriely, Phys. Lett. B 436, 10 (1998). 

\bibitem{Gor.02} S. Goriely and E. Khan, Nucl. Phys. A 706, 217 (2002).

\bibitem{Gor.04} S. Goriely, E. Khan, and M. Samyn, 
	Nucl. Phys. A 739, 331 (2004).

\bibitem{Rad.02} D. C. Radford et al., Phys. Rev. Lett. 88, 222501 (2002).

\bibitem{Ter.02} J. Terasaki, J. Engel, W. Nazarewicz, and M. Stoitsov,
	Phys. Rev. C 66, 054313 (2002).
	
\bibitem{Lei.01} A. Leistenschneider et al., Phys. Rev. Lett. 86, 5442 (2001).

\bibitem{Har.00} T. Hartmann, J. Enders, P. Mohr, K. Vogt, S. Volz, and
	A. Zilges, Phys. Rev. Lett. 85, 274 (2000).
	
\bibitem{Gov.98} K. Govaert et. al., Phys. Rev. C 57, 2229 (1998).

\bibitem{Rye.02} N. Ryezayeva et al., Phys. Rev. Lett. 89, 272502 (2002).

\bibitem{End.03} J. Enders et al., Nucl. Phys. A 724, 243 (2003).

\bibitem{Zil.02} A. Zilges, S. Volz, M. Babilon, T. Hartmann, P. Mohr,
        and K. Vogt, Phys. Lett. B 542, 43 (2002).
	
\bibitem{Suz.90} Y. Suzuki, K. Ikeda, and H. Sato,
        Prog. Theor. Phys. 83, 180 (1990).
	
\bibitem{Cha.94} J. Chambers, E. Zaremba, J.P. Adams, and  B. Castel,
        Phys. Rev. C 50, R2671 (1994).
	
\bibitem{Sag.99} H. Sagawa and T. Suzuki, Phys. Rev. C 59, 3116 (1999). 
 
\bibitem{Rei.99} P. G. Reinhard, Nucl. Phys. A 649, 305c (1999).
 
\bibitem{Col.01} G. Col\'o and P. F. Bortignon, 
	 Nucl. Phys. A 696, 427 (2001).
	 
\bibitem{Toh.01} M. Tohyama and A. S. Umar, Phys. Lett. B 516, 415 (2001).

\bibitem{Mat.01} M. Matsuo, Nucl. Phys. A 696, 371 (2001).

\bibitem{TLS.04} N. Tsoneva, H. Lenske, and Ch. Stoyanov, 
	Nucl. Phys. A 731, 273 (2004).
	
\bibitem{TLS.04a} N. Tsoneva, H. Lenske, and Ch. Stoyanov,
		Phys. Lett. B 586, 213 (2004).
	
\bibitem{Vrepyg2.01} D. Vretenar, N. Paar, P. Ring, and G. A. Lalazissis,
	Nucl. Phys. A 692, 496 (2001).
	
\bibitem{Paar.03} N. Paar, P. Ring, T. Nik\v si\' c, and D. Vretenar, 
	Phys. Rev. C 67, 034312 (2003).
	
\bibitem{Ma.01} Z. Y. Ma, N. Van Giai, A. Wandelt, D. Vretenar, and P. Ring,
	Nucl. Phys. A 686, 173 (2001).
	
\bibitem{Vre.02} D. Vretenar, N. Paar, T. Nik\v si\' c, and P. Ring,
        Phys. Rev. C 65, 021301 (2002). 
	 
\bibitem{Vre.03} D. Vretenar, N. Paar, T. Nik\v si\' c, and P. Ring,
	Phys. Rev. Lett. 91, 262502 (2003).
	
\bibitem{Nik2.02} T. Nik\v si\' c, D. Vretenar, and P. Ring,
	Phys. Rev. C 66, 064302 (2002). 
	
\bibitem{Nik1.02} T. Nik\v si\' c, D. Vretenar, P. Finelli, and P. Ring,
	Phys. Rev. C 66, 024306 (2002).
	
\bibitem{BGG.84} J.F. Berger, M. Girod, and D. Gogny,
	Nucl. Phys. A 428, 23c (1984). 
	            
\bibitem{RRM.86} P.G. Reinhard, M. Rufa, J. Maruhn, W. Greiner, and
	J. Friedrich, Z. Phys. 323, 13 (1986).
	
\bibitem{LKR.97} G.A. Lalazissis, J. K\"onig, and P. Ring,
 	Phys. Rev. C 55, 540 (1997).
	
\bibitem{CHARGE} E.G. Nadiakov, K.P. Marinova, and Yu.P. Gangrsky,
              At. Data Nucl. Data Tables 56, 133 (1994).

\bibitem{Blanc.02} F. Le Blanc et al., Eur. Phys. J. A 15, 49 (2002).
	      
\bibitem{Kra.99} A. Krasznahorkay et al., Phys. Rev. Lett. 82, 3216 (1999).

\bibitem{Ber.75} B. L. Berman and S. C. Fultz, Rev. Mod. Phys. 47, 713 (1975).

\bibitem{AW.95} G. Audi and A. H. Wapstra, Nucl. Phys. A 595, 409 (1995).


\end{thebibliography}
\end{document}